\documentclass{revtex4-2}

\usepackage[utf8]{inputenc}
\usepackage[english]{babel}

\usepackage[hidelinks]{hyperref}

\usepackage{graphicx}
\usepackage{subcaption}
\usepackage{parskip}

\usepackage{float}
\usepackage{booktabs}
\usepackage{makecell}
\usepackage{multirow}

\usepackage[figurename=Fig.]{caption}
\usepackage[tablename=Tab.]{caption}
\usepackage[font=small]{caption}

\usepackage{amsmath}
\usepackage{amssymb}
\usepackage{amsfonts}
\usepackage{mathtools}

\usepackage{upgreek}
\usepackage{xcolor}
\usepackage[normalem]{ulem}

\newcommand{\kg}[1]{{\textcolor{teal}{{#1}}}}

\begin{document}

\title{Circular Orbits and Photon Orbits at Wormhole Throats}
\date{\today}

\author{Kristian Gjorgjieski}
 \email{kristian.gjorgjieski@uol.de}
\author{Jutta Kunz}%
 \email{jutta.kunz@uni-oldenburg.de}
\affiliation{Department of Physics, Carl von Ossietzky University of Oldenburg, 26111 Oldenburg, Germany}

\author{Petya Nedkova}
 \email{pnedkova@phys.uni-sofia.bg}
\affiliation{Department of Theoretical Physics, Sofia University, Sofia 1164, Bulgaria}

\begin{abstract}
 In this work we study timelike circular orbits and photon orbits at the throat of stationary and axisymmetric wormholes. Our minimal requirements on the spacetime are the existence of a global radial coordinate $l$, which connects both sides of the wormhole, two times differentiable metric components with respect to $l$ at the wormhole throat and vanishing first derivatives of the metric components at the wormhole throat, which is the case for symmetrical wormholes. We derive expressions in terms of the metric components for the specific angular momentum $\ell$ of a test particle, that describe a possible spectrum of solutions for bound circular orbits at the wormhole throat. We identify a phase transition in the parameter space, which occurs if an ergoregion is present. Furthermore, we showcase expressions for the parameter space in terms of physical properties of the spacetime in the form of the throat circumference $C_T$, the angular frequency of the wormhole $\omega$ and the gravitational redshift $z$. An analysis of these expressions and the characteristics of the photon orbits gives constraints on wormhole properties, as it hints to possible instabilities, such as for fast rotating wormholes and wormholes with an ergoregion. Through the use of accretion disk models, the existence of possible stable circular orbits could be linked to accretion disks surrounding the wormhole throat.
\end{abstract}

\maketitle


\section{Introduction}
The theory of general relativity established a framework within which various kinds of extraordinary and compared to our day-to-day experience otherworldly concepts can exist. Among the most prominent and relevant ones are black holes. Decades after their theoretical description we now have overwhelming evidence for their existence. Moreover, they are thought to be among the dominating astrophysical bodies on larger scales, as they are believed to be the centers of galaxies. Initially established by indirect and later on by direct observations, black holes are now considered as objects within our standard models. Besides black holes, also the concept of wormholes is one of the most prominent ones within general relativity and also within theories beyond standard general relativity. The concept of wormholes was first theorized in the 1910s and 1920s \cite{Flamm,Weyl} and further developed in the 1930s, where Albert Einstein and Nathan Rosen hypothesized connections between regions of spacetime based on vacuum solutions of the Einstein field equations \cite{ERbridges}. This coined the name Einstein-Rosen bridge for such connections. Later on it was understood that their concepts corresponded to the maximal extended Schwarzschild spacetime, where such a bridge is unstable and does not represent a traversable connection. Generally speaking, a wormhole describes an object which connects two arbitrary distant regions of spacetime, where both ends of this junction can be located in the same connected spacetime or even in different ones. The latter case would correspond to a wormhole, which connects two different universes. The concepts were further developed and special conditions were discovered which wormholes should satisfy in order to ensure traversability for physical entities \cite{Fuller,Ellis, Bronnikov, Morris-Throne, Visser1989, Kanti2011jz, Burkhard, Maldacena_2021}.

Any two-way traversable wormhole should be horizonless and singularity-free. However, these conditions lead to a negative energy density needed at the throat to stabilize the connection. As hypothesized, an exotic form of matter which exhibits a negative energy density could be used to stabilize the connection. Nonetheless, there are several proposals which also extend to theories beyond general relativity, on how the need for exotic matter can be bypassed or how such objects could form naturally \cite{Cramer1995, Gravanis2007, Bronnikov_2015, Jose2021, Bakopoulos_2022}.

In order to test such theories and moreover in order to possibly identify wormholes as astrophysical bodies, questions regarding the observational features that such objects may possess need to be addressed. In certain astrophysical scenarios wormholes may possess similar observational signatures as black holes \cite{Petya2013, Gyulchev:2018, Hyat2023, Rahaman_2021, Paul_2020}. However, in other cases they exhibit distinctive phenomenology \cite{Vincent:2020, Delijski:2022, Deligianni:2021, Deligianni:2021hwt}. Revealing further possible differences compared to black holes is important for the identification of differentiating factors  for future observations.

Motivated by this, we will showcase in this work the existence of circular orbits for massive particles at the throat of wormholes for a generalized class of stationary and axisymmetric wormholes. Moreover, we will also derive expressions and conditions for the existence and stability of photon orbits. With the use of the radial effective potential we derive parameter space boundaries, which characterize a possible spectrum of circular orbit solutions. Furthermore, we showcase how this parameter space is influenced by the presence of an ergoregion and also how it depends on physical properties of the spacetime, namely the throat circumference, the angular frequency of the wormhole and the gravitational redshift. Our findings may give general hints on the stability of wormhole solutions, as for solutions with ergoregions, which are strengthened by other works \cite{ER_instability1978, ER_instability2018} and furthermore be relevant also for observational signatures in the form of accretion disks and matter accumulations, as these are usually modeled on the presence of stable circular orbits.

This work is structured as follows. In Section 2 we first derive limits for the range of existence and boundedness of circular orbits at the throat with regard to the specific angular momentum of a test particle. Section 3 expands this analysis on the stability of the orbits. In Section 4 we derive analogously expressions for photon orbits and conditions for their instability. Afterwards in Section 5 we concretize our assumptions on the wormhole and assume an asymptotically flat wormhole, where both sides are symmetrical to each other. From a generalized form of the metric we derive then expressions for the parameter space boundaries in terms of physical properties of the spacetime. We use the metric signature $(-1,1,1,1)$ and geometrized units where $G = c = 1$, with $G$ as the gravitational constant and $c$ as the speed of light.


\section{Circular orbits at Wormhole Throats}

We assume here a stationary and axially symmetric wormhole, where both sides of the wormhole share locally the same properties with respect to the wormhole throat, i. e. at least a local mirror symmetry, which leads to vanishing first derivatives of the metric components. Furthermore, we suppose that both sides are connectable by a global radial coordinate $l \in (-\infty,\infty)$, with $l = 0$ designating the wormhole throat. We define the region $l \in (-\infty, 0]$ as the lower side and $l \in [0, \infty)$ as the upper side of the wormhole. This connection should be definable in such a way, that the metric components $g_{\mu\nu}$ are at least $\mathcal{C}^2$ functions with respect to $l$ at the wormhole throat. These constraints are trivially satisfied for a smoothly connected symmetrical wormhole. However, it should be noted that a number of known rotating wormhole solutions do not satisfy these constraints \cite{Chew_2016, Chew_2018, Deligianni_2021}. Nonetheless, there also exist known numerical as well as analytical rotating wormhole geometries, which do satisfy these conditions \cite{Chew_2019, Mazza_2021, Teo}. The squared line element of such wormhole metric can be written in generalized spherical coordinates as

\begin{align}
    ds^2 = g_{tt} dt^2 + g_{ll} dl^2 + g_{\theta \theta} d\theta^2 + g_{\varphi \varphi} d \varphi^2 + 2 g_{t \varphi} dt d\varphi.
\end{align}

Circular motion of particles is characterized by a vanishing radial velocity component. We restrict ourselves here to planar circular motion within the equatorial plane ($\theta = \pi/2$ and $d\theta = 0$). The Lagrangian $\mathcal{L}^*$ describing the motion for massive particles can then be written as

\begin{align}
    \mathcal{L}^* = g_{ll} \dot{l}^2 + g_{tt} \dot{t}^2 + g_{\varphi \varphi} \dot{\varphi}^2 + 2 g_{t \varphi} \dot{t} \dot{\varphi} = -1, \label{eq:Lagrangian}
\end{align}

where the dot marks the derivative with respect to the proper time $\tau$. The coordinates $t$ and $\varphi$ are due to the stationarity and axial symmetry of the spacetime cyclic coordinates, which lead to the presence of two constants of motion, namely the energy $E$ and the angular momentum $L$ of the particle. The derivatives $\dot{t}$ and $\dot{\varphi}$ can then be expressed through $E$ and $L$ as

\begin{align}
    \dot{t} = \frac{g_{\varphi \varphi}E + g_{t\varphi}L}{g_{t \varphi}^2 - g_{tt} g_{\varphi \varphi}} \ \ \ \ ; \ \ \ \ \dot{\varphi} = -\frac{g_{tt} L + g_{t\varphi} E}{g_{t \varphi}^2 - g_{tt} g_{\varphi \varphi}},
\end{align}

inserting these expressions into Eq. (\ref{eq:Lagrangian}) leads to

\begin{align}
    g_{ll}\dot{l}^2 - \frac{1}{g_{t\varphi}^2 - g_{\varphi\varphi} g_{tt}} \left(g_{\varphi\varphi} E^2  +  g_{tt}L^2 + 2 g_{t \varphi} E L \right) + 1 = g_{ll} \dot{l}^2 + V_{eff} = 0.
\end{align}

Circular motion requires the conditions $V_{eff} = 0$ and $\partial_l V_{eff} = 0$. This system of two equations can be solved for the energy $E$ and the angular momentum $L$ of a particle on a circular orbit,

\begin{align}
    V_{eff} = 0 \ \Leftrightarrow \ \ \ \ \ \ \ \ \ \ \ \ \ \ g_{\varphi\varphi} E^2 + g_{tt}L^2 + 2 g_{t \varphi} E L - (g_{t\varphi}^2 - g_{\varphi \varphi} g_{tt}) &= 0 \label{eq:V} \\
     \partial_l V_{eff} = 0 \ \Leftrightarrow \ \ \ \  E^2 \partial_l g_{\varphi \varphi} + L^2 \partial_l g_{tt} + 2 E L \partial_l g_{t \varphi} - \partial_l (g_{t\varphi}^2 - g_{tt} g_{\varphi \varphi}) &= 0. \label{eq:ParialV}
\end{align}

In view of a simple notation, from now on every quantity will denote its value at the throat if not stated otherwise. Since the first derivatives of the metric components vanish at the wormhole throat, Eq. (\ref{eq:ParialV}) is always trivially satisfied. As a consequence the system of equations is under-determined, and a spectrum of solutions for $E$ and $L$ is possible. To analyze the spectrum, Eq. (\ref{eq:V}) can be rearranged for the energy $E$,

\begin{align}
    \frac{g_{t \varphi}^2 - g_{tt} g_{\varphi \varphi}}{g_{tt} \ell^2 + 2g_{t \varphi} \ell + g_{\varphi \varphi}} = E^2 \label{eq:Energy_ell},
\end{align}

where $\ell = \frac{L}{E}$ is the specific angular momentum of the particle.  It follows immediately that $E^2 > 0$ is the boundary for the existence of circular orbits. Furthermore, for a normalized mass-energy, bound circular orbits require that the energy of the particle must be $E \leq 1$, with $E = 1$ corresponding to the marginally bound orbit. Combining these boundaries, leads to a possible spectrum of solutions for $\ell$, for which bound circular orbits exist at the wormhole throat. The Eq. (\ref{eq:Energy_ell}) can then be written as an inequality for the parameter space of $\ell$, for which the orbits exist,

\begin{align}
    0 <\frac{g_{t \varphi}^2 - g_{tt} g_{\varphi \varphi}}{g_{tt} \ell^2 + 2g_{t \varphi} \ell + g_{\varphi \varphi}} \leq 1. \label{eq:CO_inequality}
\end{align}

Solving for the limits of this inequality gives the limiting values for $\ell$ and furthermore defines the shape of the parameter space. The numerator in the expression corresponds to the negative determinant of the two-dimensional induced metric for $g_{ij}$ with $(i,j) = t,\varphi$, which is always negative. The numerator is therefore always positive. As a consequence the denominator in (\ref{eq:CO_inequality}) must strictly be positive in order to satisfy the lower bound of the inequality. Thus, the roots of the denominator give the boundary for the existence of circular orbits, 

\begin{align}
    g_{tt} \ell^2 + 2 g_{t\varphi} \ell + g_{\varphi \varphi} > 0 \ \ \ \Rightarrow \ \ \ell_\pm  = \frac{-g_{t\varphi} \pm \sqrt{g_{t \varphi}^2 - g_{tt} g_{\varphi \varphi}}}{g_{tt}} \coloneqq \ell^\pm_E \label{eq:ell_E},
\end{align}

where we define the boundary of existence for circular orbits as $\ell_E^\pm$. In order to determine the region for the bound circular orbits, the expression $(\ref{eq:CO_inequality})$ can be solved for the equality in the upper limit, 

\begin{align}
    \frac{g_{t \varphi}^2 - g_{tt} g_{\varphi \varphi}}{g_{tt} \ell^2 + 2g_{t \varphi} \ell + g_{\varphi \varphi}} &\leq 1 \ \ \Rightarrow \ \ \ell_\pm =  \frac{- g_{t \varphi} \pm \sqrt{ \left(g^2_{t \varphi} - g_{tt} g_{\varphi \varphi} \right) \left( 1 + g_{tt} \right)}}{g_{tt}} \coloneqq \ell^\pm_B \label{eq:ell_B}
\end{align}

where we define $\ell^\pm_B$ as the boundary for closed circular orbits. For both limits, we assumed that $g_{tt} \neq 0$ at the throat. In the special case of $g_{tt} = 0$ the boundaries are given by $\ell_E = -\frac{g_{\varphi \varphi}}{2 g_{t\varphi}}$ and $\ell_D = \frac{g_{t \varphi}^2- g_{\varphi \varphi}}{2 g_{t \varphi}}$. Furthermore, it follows from Eq. (\ref{eq:ell_B}) that for the existence of bound orbits $g_{tt} > -1$ must hold at the throat. To determine which of the boundaries is the lower bound and which is the upper bound, the differences $\ell_E^+ - \ell_E^-$ and $\ell_B^+ - \ell_B^-$ can be analyzed,

\begin{align}
    \ell^+_E - \ell^-_E = 2\frac{\sqrt{g_{t \varphi}^2 - g_{tt} g_{\varphi \varphi}}}{g_{tt}} \ \ \ ; \ \ \ \ell^+_B - \ell^-_B = 2\frac{\sqrt{(g_{t \varphi}^2 - g_{tt} g_{\varphi \varphi})(1 + g_{tt})}}{g_{tt}}. \label{eq:differences}
\end{align}

Since both inequalities are quadratic expressions of $\ell$, they possess an extremum between the boundaries. With consideration of the sign of the inequalities, the second derivative of the inequalities with respect to $\ell$ determines then if the allowed values for $\ell$ lie in between the boundaries or outside of them. Furthermore, the relation between the boundaries of existence and the boundaries for bound circular orbits can be analyzed by taking the differences $\ell_E^+ - \ell_B^+$ and $\ell_E^- - \ell_B^-$,

\begin{align}
    \ell_E^+ - \ell_B^+ &= \frac{\sqrt{g_{t \varphi}^2 - g_{tt} g_{\varphi \varphi}} - \sqrt{(g_{t\varphi}^2 - g_{tt} g_{\varphi \varphi})(1 + g_{tt})}}{g_{tt}} \label{eq:difference_E_B_p}\\
    \ell_E^- - \ell_B^- &= \frac{\sqrt{(g_{t\varphi}^2 - g_{tt} g_{\varphi \varphi})(1 + g_{tt})} - \sqrt{g_{t \varphi}^2 - g_{tt} g_{\varphi \varphi}}}{g_{tt}}. \label{eq:difference_E_B_n}
\end{align}

The second derivatives of the inequalities (with respect to $\ell$) and each of the defined differences depend all on the sign of $g_{tt}$ at the throat. We differentiate here now between both cases $g_{tt} < 0$ and $g_{tt} > 0$ (which usually corresponds to an ergoregion) and analyze their implications.

If $g_{tt} < 0$, then the expressions in (\ref{eq:differences}) give $\ell_E^+ <\ell_E^-$ and $\ell_B^+ < \ell_B^-$ as the numerator is always positive. Furthermore, the second derivative of (\ref{eq:ell_E}) is negative, which implies that the extremum between the roots is a maximum. As a consequence the allowed region for $\ell$ lies in between $\ell_E^+$ and $\ell_E^-$. The second derivative of (\ref{eq:ell_B}) is positive and the allowed region for $\ell$ lies also in between $\ell_B^+$ and $\ell_B^-$. The difference in (\ref{eq:difference_E_B_p}) is negative as the numerator is positive, since the expression in the second square root is smaller than in the first square root, due to $(1 + g_{tt}) \in (0,1)$. Following the same reasoning the difference in (\ref{eq:difference_E_B_n}) is always positive. We have therefore $\ell_E^+ < \ell_E^-$ and $\ell_B^- < \ell_E^-$. Combining all inequalities leads to $\ell_E^+ < \ell_B^+ < \ell_B^- < \ell_E^-$. The parameter space for bound circular orbits is thus given by $\ell \in [\ell_B^+,\ell_B^-]$.

If $g_{tt} > 0$, then the expressions in (\ref{eq:differences}) give $\ell_E^- < \ell_E^+$ and $\ell_B^- < \ell_B^+$. The second derivative of (\ref{eq:ell_E}) is positive and the extremum between the roots is a minimum. The region for allowed values of $\ell$ lies outside the interval $[\ell_E^-,\ell_E^+]$. The second derivative of (\ref{eq:ell_B}) is negative and the extremum is a maximum, thus the allowed region of $\ell$ for bound orbits lies also outside the interval $(\ell_B^-,\ell_B^+)$. The differences in (\ref{eq:difference_E_B_p}) and (\ref{eq:difference_E_B_n}) lead to $\ell_E^+ < \ell_B^+$ and $\ell_B^- < \ell_E^-$, since now $(1 + g_{tt}) > 1$. Bringing all inequalities together leads to $\ell_B^- < \ell_E^- < \ell_E^+ < \ell_B^+$. The parameter space for bound circular orbits is thus given by $\ell \in (-\infty, \ell_B^-] \ \vee \ \ell \in [\ell_B^+, \infty)$. Following Tab. \ref{tab:boundaries} gives a compact overview of the behavior of the boundaries.

\begin{table}[H]
\centering
\vline
\begin{tabular}{c|c|c|c}
     \toprule
       & \text{Existence} & \text{Boundedness} & \text{Parameter Space} \\[1mm]
     \midrule
     $g_{tt} < 0 $ & $\ell_E^+ < \ell_E^-$ & $\ell_B^+ < \ell_B^-$ & $\ell \in [\ell_B^+,\ell_B^-]$ \\[2mm] \midrule
     $g_{tt} > 0$ & $\ell_E^- < \ell_E^+$ & $\ell_B^- < \ell_B^+$ & $\ell \in (-\infty, \ell_B^-] \ \vee \ \ell \in [\ell_B^+, \infty)$ \\[2mm]
     \bottomrule
\end{tabular}
\vline
\caption{Relationship between the boundaries of existence and boundedness for circular orbits at the throat, as well as the corresponding range for the parameter $\ell$ for which they exist. A different sign of $g_{tt}$ at the throat leads to a phase transition in the parameter space. It should be noted that bound orbits require furthermore $g_{tt} > -1$ as a necessary condition in case of $g_{tt} < 0$.}
\label{tab:boundaries}
\end{table}


\section{Stability of orbits}

Considering now the stability of the orbits, stable circular motion requires the extremum of $V_{eff}$ to be a potential minimum and thus the inequality $\partial_l^2 V_{eff} \geq 0$ has to be satisfied at the throat,

\begin{align}
     \partial_l^2 V_{eff} \geq 0 \ \Leftrightarrow \ \  E^2 \partial_l^2 g_{\varphi \varphi} + L^2 \partial_l^2 g_{tt} + 2 E L \partial_l^2 g_{t \varphi} - \partial_l^2 (g_{t\varphi}^2 - g_{tt} g_{\varphi \varphi}) \leq 0. \label{eq:partialV2}
\end{align}

Rearranging this inequality in terms of $\ell$ and furthermore inserting the expression $(\ref{eq:Energy_ell})$ for the squared energy $E^2$ leads to

\begin{align}
    \ell^2 \partial_l^2 g_{tt} + 2 \ell \partial_l^2 g_{t \varphi} + \partial_l^2 g_{\varphi \varphi} - \frac{\partial_l^2(g_{t\varphi}^2 - g_{tt} g_{\varphi \varphi})}{g_{t \varphi^2} - g_{tt} g_{\varphi \varphi}}(\ell^2 g_{tt} + 2 \ell g_{t \varphi} + g_{\varphi \varphi}) \leq 0. 
\end{align}

This represents again a quadratic inequality for the specific angular momentum $\ell$, where the boundaries of stability are given by solving for the equality. For the sake of more compact expressions we denote here $D = g_{t \varphi}^2 - g_{tt} g_{\varphi \varphi}$, the inequality and the boundaries are then given by

\begin{align}
    & \ell^2 \left(D\partial_l^2 g_{tt} - g_{tt} \partial_l^2 D \right) + 2 \ell \left( D\partial_l^2 g_{t \varphi} - g_{t\varphi} \partial_l^2 D \right) + D\partial_l^2 g_{\varphi \varphi} - g_{\varphi \varphi}\partial_l^2 D \leq 0  \label{eq:ell_S} \\
    \Rightarrow \ & \ell_\pm = \frac{D \partial_l^2 g_{t \varphi} - g_{t \varphi}\partial_l^2 D \pm \sqrt{(g_{t \varphi} \partial_l^2 D - D \partial_l^2 g_{t \varphi})^2 - (g_{tt} \partial_l^2 D - D \partial_l^2 g_{tt})(g_{\varphi \varphi} \partial_l^2 D - D \partial_l^2 g_{\varphi \varphi})}}{g_{tt} \partial_l^2 D - D \partial_l^2 g_{tt}} \coloneqq \ell_S^\pm,
\end{align}

where we define the boundaries of the stability for circular orbits as $\ell_S^\pm$. Stable orbits only exist if the expression under the square root is not negative, which gives a necessary condition for stability,

\begin{align}
    (g_{t \varphi} \partial_l^2 D - D \partial_l^2 g_{t \varphi})^2 - (g_{tt} \partial_l^2 D - D \partial_l^2 g_{tt})(g_{\varphi \varphi} \partial_l^2 D - D \partial_l^2 g_{\varphi \varphi}) \geq 0.
\end{align}

In order to analyze the relation between the boundaries, we conduct the same procedure as for the limits on existence and boundedness in Section 2. The difference $\ell_S^+ - \ell_S^-$ and the second derivative with respect to $\ell$ of the inequality in $(\ref{eq:ell_S})$ are given by

\begin{align}
    &\ell_S^+ - \ell_S^- = 2\frac{\sqrt{(g_{t \varphi} \partial_l^2 D - D \partial_l^2 g_{t \varphi})^2 - (g_{tt} \partial_l^2 D - D \partial_l^2 g_{tt})(g_{\varphi \varphi} \partial_l^2 D - D \partial_l^2 g_{\varphi \varphi})}}{g_{tt} \partial_l^2 D - D \partial_l^2 g_{tt}} \\
    &\partial_\ell^2(\ell^2 \left(D\partial_l^2 g_{tt} - g_{tt} \partial_l^2 D \right) + 2 \ell \left( D\partial_l^2 g_{t \varphi} - g_{t\varphi} \partial_l^2 D \right) + D\partial_l^2 g_{\varphi \varphi} - g_{\varphi \varphi}\partial_l^2 D) = 2\left(D\partial_l^2 g_{tt} - g_{tt} \partial_l^2 D \right).
\end{align}

It follows from these expressions, that the stability behavior is determined by the sign of $D\partial_l^2 g_{tt} - g_{tt} \partial_l^2 D$ at the throat. In general it is not feasible to evaluate this expression, since it depends not only on the sign of $g_{tt}$ and $\partial_l^2 g_{tt}$, but also on the other metric components and their second derivatives, as well as on their relationship between each other. However, if the expression is positive at the throat, then $\ell_S^+ < \ell_S^-$ and the extremum between the roots $\ell_S^\pm$ is a minimum, the region of stable orbits is then given by $\ell \in [\ell_S^+, \ell_S^-]$. If the expression is negative at the throat, then $\ell_S^- < \ell_S^+$ and the extremum between the roots is a maximum, the region of stable orbits is then given by $\ell \in (-\infty,\ell_S^-] \ \vee \ \ell \in [\ell_S^+,\infty)$.


\section{Photon Orbits at wormhole throats}

Photon orbits describe circular orbits for photons or in general for massless particles. We follow here the same procedure as in Section 2 and start with the Lagrangian $\mathcal{L}^*$ for massless particles. In case of planar motion within the equatorial plane it is now given by

\begin{align}
    \mathcal{L}^* = g_{ll} \dot{l}^2 + g_{tt} \dot{t}^2 + g_{\varphi \varphi} \dot{\varphi}^2 + 2 g_{t \varphi} \dot{t} \dot{\varphi} = 0.
\end{align}

This leads to a slightly different radial effective potential $V_{eff}^{P} = V_{eff} - 1$. The impact parameter $\ell = \frac{L}{E}$ of photons corresponding to the photon orbits is determined by the roots of the radial effective potential $V_{eff}^{P}$, which gives rise to 

\begin{align}
    V^{P}_{eff} = 0 \ \Leftrightarrow \ g_{tt} \ell^2 + 2 \ell g_{t \varphi} + g_{\varphi \varphi} = 0.
\end{align}

This equation is identical to the equation for the limits of existence of circular orbits for particles with mass. The impact parameters of the photon orbits, $\ell_P^\pm$, are thus identical to the boundaries of existence of circular orbits for particles with mass, $\ell_P^\pm = \ell_E^\pm$. For the static limit, $g_{t\varphi} = 0$, the expressions for the impact parameter at the throat are identical to the ones presented in \cite{Tsukamoto_2024}. The second derivative of $V_{eff}^{P}$ determines the stability of the photon orbits, for stable orbits it can not be negative,

\begin{align}
    \partial_l^2 V_{eff}^{P} \geq 0 \ \Leftrightarrow & \ \ell^2 \partial_l^2 g_{tt} + 2 \ell \partial_l^2 g_{t \varphi} + \partial_l^2 g_{\varphi \varphi} \leq 0 \\
    \Rightarrow & \ \ell_\pm = \frac{- \partial_l^2 g_{t \varphi} \pm \sqrt{(\partial_l^2 g_{t \varphi})^2 - \partial_l^2 g_{tt} \partial_l^2 g_{\varphi \varphi}}}{\partial_l^2 g_{tt}} \coloneqq P_S^\pm, \label{eq:P_S}
\end{align}

where we define $P_S^\pm$ as the critical value for the stability of the photon orbits. From $P_S^\pm$ it follows directly, that the stability behavior is determined by the second derivative of $g_{tt}$ at the throat. If $\partial_l^2 g_{tt} > 0$, then $P_S^+ > P_S^-$ and the photon orbits are stable if $\ell_P^\pm \in [P_S^-,P_S^+]$. If $\partial_l^2 g_{tt} < 0$, then $P_S^+  < P_S^-$ and the stable region lies outside the interval $(P_S^+,P_S^-)$, stability requires therefore $\ell_P^\pm \notin (P_S^+,P_S^-)$.

Stable photon orbits could hint to instabilities of the spacetime \cite{Cunha, Xavier}. The general stability of the wormhole spacetime may thus possibly be analyzed by evaluating the expressions for stable and unstable photon orbits. The conditions for unstable orbits correspond to an interchanged relation of $\ell_P^\pm$ with regard to the stability regions bounded by $P_S^\pm$. For both cases with regard to the sign of $\partial_l^2 g_{tt}$ the inequalities are the same, writing them out by inserting the expressions for $\ell_P^\pm$ and $P_S^\pm$ leads to the following constraints on the metric functions,

\begin{align}
    \frac{-g_{t\varphi} \pm \sqrt{g_{t\varphi}^2 - g_{tt} g_{\varphi \varphi}}}{g_{tt}} &<  \frac{- \partial_l^2 g_{t \varphi} - \sqrt{(\partial_l^2 g_{t \varphi})^2 - \partial_l^2 g_{tt} \partial_l^2 g_{\varphi \varphi}}}{\partial_l^2 g_{tt}} \label{eq:Unstable_PO} \\ 
    \vee \ \ \frac{-g_{t\varphi} \pm \sqrt{g_{t\varphi}^2 - g_{tt} g_{\varphi \varphi}}}{g_{tt}} &> \frac{- \partial_l^2 g_{t \varphi} + \sqrt{(\partial_l^2 g_{t \varphi})^2 - \partial_l^2 g_{tt} \partial_l^2 g_{\varphi \varphi}}}{\partial_l^2 g_{tt}}, \\
\end{align}

where for $\partial_l^2 g_{tt} > 0$ either the upper inequality or the lower inequality must be satisfied for unstable photon orbits and for $\partial_l^2 g_{tt} < 0$ both inequalities have to be satisfied.


\section{Boundaries of the parameter space in terms of physical quantities}
In order to conduct a deeper investigation of the calculated bounds for the parameter space, we assume now an asymptotically flat wormhole, where the upper and lower side are symmetric to each other. Furthermore, we assume that the squared line element can be written in terms of the metric functions as follows,

\begin{align}
    ds^2 = -Ndt^2 + R dl^2 + T d\theta^2 + K \left(d\varphi - \omega dt \right)^2,
\end{align}

where the functions $N, \ R,\ T, \ K$ and $\omega$ possess axial symmetry and are stationary, they depend therefore only on the global radial coordinate $l$ and on the polar coordinate $\theta$. The function $\omega$ characterizes the rotation of the spacetime and is strictly increasing in terms of the angular momentum $J$ of the wormhole. At the wormhole throat it corresponds to the angular frequency of the wormhole. Due to the asymptotic flatness we have $R \rightarrow 1$ and $\omega \rightarrow 0$ for $|l| \rightarrow \infty$, which also leads to $N \rightarrow 1$ for $|l| \rightarrow \infty$. The metric functions $T$ and $K$ are monotonically increasing \kg{for larger values of $|l|$}, with $T,N \rightarrow l^2$ in the equatorial plane for $|l| \gg 0$. The metric components $g_{tt}$, $ g_{t \varphi}$ and $g_{\varphi \varphi}$ are given in terms of the metric functions by

\begin{align}
    g_{tt} = K \omega^2 - N \ \ \ ; \ \ \ g_{t \varphi} = - K \omega \ \ \ ; \ \ \ g_{\varphi \varphi} = K. \label{eq:metric_components}
\end{align}

Inserting these expressions into the boundaries $\ell_E^\pm$ and $\ell_B^\pm$ leads to

\begin{align}
    \ell_E^\pm = \frac{K \omega \pm \sqrt{NK}}{Kw^2 -N} = \ell_P^\pm \ \ \ ; \ \ \ 
    \ell_B^\pm = \frac{K \omega \pm \sqrt{NK(1 + K\omega^2 - N)}}{Kw^2 -N}.
\end{align}

To incorporate physical properties of the spacetime into these expressions, the metric function $K$ can be expressed through the circumference of the wormhole throat $C_T$,

\begin{align}
    C_T = \int_0^{2 \pi} \sqrt{g_{\varphi \varphi}} d\varphi = 2\pi \sqrt{K} \ \ \ \Leftrightarrow \ \ \ K = \frac{C_T^2}{4 \pi^2}.
\end{align}

Furthermore, if there is no ergoregion present around the wormhole ($g_{tt} < 0$), the metric function $N$ can be expressed in terms of the gravitational redshift $z$, which a distant observer would measure for signals coming from a static source at the wormhole throat,

\begin{align}
    1 + z = (-g_{tt})^{-\frac{1}{2}} = (N - K\omega^2)^{-\frac{1}{2}} = \left(N - \frac{C_T^2 \omega^2}{4 \pi^2} \right)^{-\frac{1}{2}} \ \ \ \Leftrightarrow \ \ \ N = \frac{4\pi^2 + C_T^2\omega^2 (1 +z)^2}{4 \pi^2 (1+z)^2}.
\end{align}

In order to investigate how the parameter space for bound circular orbits varies with respect to the physical properties of the spacetime, the interval between the boundaries $\ell_B^\pm$ can be analyzed. The interval size $\mathcal{B}$ is given by the absolute value of the difference between the boundaries. In case of no ergoregion, the interval size is identical to the parameter space volume for bound circular orbits and it is fully expressible through physical properties of the spacetime in the form of $C_T$, $\omega$ and $z$,

\begin{align}
    \mathcal{B} = |\ell_B^\pm - \ell_B^\mp| &= 2\frac{\sqrt{NK(1 + K \omega^2 -N)}}{|K \omega^2 -N|} \\
                                         \Leftrightarrow \  \mathcal{B}(C_T,\omega,z)   &= \frac{C_T}{2\pi^2} \sqrt{\left( 4 \pi^2 + C_T^2 \omega^2 (1+z)^2 \right)(z^2 + 2z)}. 
\end{align}

For the absence of an ergoregion, the parameter space for bound circular orbits is a monotonically increasing function of these physical spacetime properties. A greater throat circumference $C_T$ and a greater angular frequency $\omega$ of the wormhole, as well as a higher redshift factor $z$ from a static source at the throat, would all correspond to the growth of the parameter space for bound circular orbits. Orbits are thus more likely to occur for faster rotating wormholes than for slower rotating ones, as well as for bigger than smaller ones and also for more compact ones as indicated by the dependence on the redshift factor. Analogously, the impact parameter of the photon orbits can be also expressed through $C_T$, $\omega$ and $z$,

\begin{align}
    \ell_P^\pm(C_T,\omega,z) = -\frac{C_T (1 + z)}{4 \pi^2} \left( C_T \omega (1+z) \pm \sqrt{4 \pi^2 + C_T^2 \omega^2 (1 + z)^2} \right).
\end{align}

If an ergoregion is present, the metric function $N$ can not be expressed solely through $C_T$, $\omega$ and $z$. A source located at the wormhole throat can not be static and must be in motion, which leads to inseparable expressions for the redshift $z$ with regard to $N$. Trying to incorporate the redshift factor in an expression for $\mathcal{B}$ would thus only complicate the expression. For the analysis of wormholes with an ergoregion we therefore express $\mathcal{B}$ as a function of $C_T$, $\omega$ and $N$,

\begin{align}
    \mathcal{B}(C_T,\omega,N) = 2 C_T \frac{\sqrt{4 \pi^2 N (1 - N) + NC_T^2 \omega^2}}{C_T^2 \omega^2 - 4 \pi^2 N},
\end{align}

where the denominator is always positive since $g_{tt} > 0$. Due to the transition of the parameter space behavior for orbits if $g_{tt} > 0$ at the throat, the interval size $\mathcal{B}$ now represents the size of the region outside of the parameter space for bound circular orbits. Considering a variation of the angular frequency for fixed $C_T$ and $N$, the interval size $\mathcal{B}$ is shrinking for faster rotating wormholes, with $\mathcal{B}(\omega) = \mathcal{O}\left(\frac{1}{\omega}\right)$. Thus, for an increasing $\omega$ the interval size approaches asymptotically zero and the parameter space for bound circular orbits approaches asymptotically $\ell \in (-\infty, \infty)$. This could be interpreted as a hint to possible instabilities of wormholes with ergoregions and faster rotating wormholes, as more and more mass could accumulate at the throat. Fixing $\omega$ and $N$ and increasing the throat circumference $C_T$ leads also to a decrease of $\mathcal{B}$ as the numerator gets smaller and the denominator bigger. For growing $C_T$ it approaches asymptotically $2\frac{\sqrt{N}}{\omega}$. General statements on the behavior of $\mathcal{B}$ with regard to $N$ are difficult, since $N$ is coupled to the other quantities by $\frac{C_T \omega^2}{ 4 \pi^2} > N$ which must be satisfied in case of an ergoregion. Regarding the impact parameters of photon orbits, they can be analogously written as

\begin{align}
    \ell_P^\pm(C_T,\omega,N) = \frac{C_T^2 \omega \pm 2 \pi C_T \sqrt{N}}{C_T^2 \omega^2 - 4 \pi^2 N}.
\end{align}

Considering now the stability, the expressions for the metric components (Eqs. (\ref{eq:metric_components})) can be inserted into the Eq. (\ref{eq:ell_S}) for the boundaries $\ell_S^\pm$ and into equation (\ref{eq:P_S}) for the boundaries $P_S^\pm$,

\begin{align}
    \ell_S^\pm &= \frac{K^2 \omega \partial_l^2 N - N K^2 \partial_l^2 \omega \pm \sqrt{K^4 \omega^2 (\partial_l^2 N)^2 + K^2 N^2 \partial_l^2 N \partial_l^2 K}}{K^2 \omega^2 \partial_l^2 N - N^2 \partial_l^2 K - 2 \omega N K^2 \partial_l^2 \omega} \\
    P_S^\pm &= \frac{K\partial_l^2 \omega + \omega \partial_l^2 K \pm \sqrt{K^2 (\partial_l^2 \omega)^2 + \partial_l^2 N \partial_l^2 K}}{2K\omega \partial_l^2 \omega + \omega^2 \partial_l^2 K - \partial_l^2 N}.
\end{align}

Due to the complexity of these expressions and the various possible relationships between the metric functions and their derivatives, it is not feasible to derive general statements out of these expressions. However, by considering more specific cases a deeper analysis can be conducted. In case of a static wormhole ($\omega = 0$), the expressions simplify to

\begin{align}
    \ell_S^\pm = \mp \frac{K}{N} \sqrt{\frac{\partial_l^2 N}{\partial_l^2 K}} \ \ \ ; \ \ \ P_S^\pm = \mp \sqrt{\frac{\partial_l^2K}{\partial_l^2N}},
\end{align}

from which follows that $\partial_l^2 N$ and $\partial_l^2 K$ need to have the same sign for the existence of stable circular orbits. Applying the inequalities from expression (\ref{eq:Unstable_PO}) for the instability of the photon orbits leads to the constraint

\begin{align}
    \sqrt{\frac{K}{N}} < \sqrt{\frac{\partial_l^2K}{\partial_l^2N}} \ \Leftrightarrow \ \sqrt{\frac{K \partial_l^2 N}{N \partial_l^2 K}} < 1, \label{eq:unstable_PO_static_WH}
\end{align}

which needs to be fulfilled by a static wormhole solution in order to avoid stable photon orbits on the throat. Using this relationship in the expression for $\ell_S^\pm$ and taking the expressions for $\ell_E^\pm = \mp \sqrt{\frac{K}{N}}$ as well as $\ell_B^\pm = \mp \sqrt{\frac{K(1-N)}{N}}$ into account, gives rise to

\begin{align}
    |\ell_S^\pm| = \sqrt{\frac{K}{N}} \sqrt{\frac{K \partial_l^2 N}{N \partial_l^2 K}} < |\ell_E^\pm| \ \ \ ; \ \ \ |\ell_B^\pm| = \sqrt{\frac{K}{N}} \sqrt{1-N} < |\ell_E^\pm|. \label{eq:inequalites_absV}
\end{align}

If instead the photon orbits are stable, then the inequality in (\ref{eq:unstable_PO_static_WH}) is reversed and $|\ell_S^\pm| > |\ell_E^\pm|$. Assuming that $K$ is minimal at the throat, the parameter space region for stable orbits is located inside the interval $(\ell_S^+,\ell_S^-)$. Thus, in case of stable photon orbits also the whole spectrum of circular orbits for massive particles  is stable. However, even for unstable photon orbits the whole spectrum of circular orbits could be stable, namely if the inequality $|\ell_B^\pm| < |\ell_S^\pm| < |\ell_E^\pm|$ holds. From the expressions in (\ref{eq:inequalites_absV}) this inequality leads to the following constraints on the metric functions,

\begin{align}
    1 - N < \frac{K \partial_l^2 N}{N \partial_l^2 K} \ \Leftrightarrow 0 < \frac{\partial_l^2 N}{N} + \frac{\partial_l^2 K}{K} (N-1),
\end{align}

where the term on the right is always positive as $N \in (0,1)$ and $\partial_l^2 K > 0$ as $K$ is assumed to be minimal at the throat. If $\partial_l^2 N > 0$, then the inequality is always trivially satisfied and despite unstable photon orbits, all bound circular orbits for massive particles are stable. However, due to the asymptotic flatness of the spacetime, the metric function $N$ can only be minimal at the throat if $g_{tt}$ possesses at least one outer local maximum. When it comes to orbital properties of the spacetime, local maxima of $g_{tt}$ can be linked to the existence of static orbits \cite{StaticOrbits}. Thus, if the wormhole spacetime exhibits a static orbit, then all circular orbits at the wormhole throat are stable. This relationship symbolizes the extent to which the properties of the spacetime geometry far from the throat also influence the characteristics and properties of motion at the throat.


\section{Conclusion and Outlook}
In this work we derived expressions for stationary and axisymmetric wormholes, which describe circular orbits and photon orbits at the wormhole throat. These expressions depend solely on the metric components, and they can be written as ranges with regard to the specific angular momentum $\ell$, for which bound circular orbits exist at the throat. We considered a generalized class of wormholes, where one side can be connected to the other side by a global radial coordinate $l$ with at least two times differentiable metric components at the wormhole throat. We furthermore considered solutions that possess at least locally a mirror symmetry around the throat, which leads to vanishing first derivatives of the metric components at the throat.

Under these assumptions we derived generalized expressions for the limits of the range of existence, $\ell_E^\pm$, for the timelike circular orbits, which are identical to the impact parameter of the photon orbits $\ell_P^\pm$. Furthermore, we derived expressions for the boundaries of bound circular orbits $\ell_B^\pm$ where we found the inequality $g_{tt} > -1$ at the throat as a necessary condition for the existence of bound circular orbits. The parameter space for the existence and boundedness of orbits mainly depends on the sign of $g_{tt}$ at the throat. In the presence of an ergoregion the parameter space undergoes a phase transition where the spectrum of solutions lies outside the characteristic range for solutions without an ergoregion and extends to infinity. This could hint at instabilities for wormholes with ergoregions, as only a small interval of values for $\ell$ prohibits circular orbits at the throat.

Considering the stability of the orbits, it is difficult to derive general statements from the expression for the boundaries of the stability $\ell_S^\pm$ due to its complexity. However, with regard to the stability of the photon orbits it is possible to derive constraints in the form of inequalities containing the metric components and their second derivatives, that the spacetime has to fulfill in order to avoid stable photon orbits, which are in general linked to instabilities of the spacetime.

Narrowing the generalizations of the spacetime further down, and assuming a symmetrical and asymptotically flat wormhole, we showed that the regions of existence and stability of the circular orbits can be expressed through physical properties of the spacetime. Here we differentiated between the absence and presence of an ergoregion. If there is no ergoregion present, the parameter space $\mathcal{B}$ for bound circular orbits can be fully expressed through the throat circumference $C_T$, the angular frequency of the wormhole $\omega$ and the redshift $z$ of a static source at the throat measured by a distant observer. The parameter space $\mathcal{B}$ is a strictly increasing function of all mentioned physical properties, thus faster rotating or bigger wormholes, as well as more compact ones, exhibit a greater parameter spectrum for circular orbits at their throat. If an ergoregion is present, the interval size $\mathcal{B}$ represents due to the phase shift the region outside the parameter space for bound circular orbits. We showed that it can be expressed as a function of $C_T$, $\omega$ and the value of the metric function $N$ at the throat. In contrast to the absence of an ergoregion, $\mathcal{B}$ is now a decreasing function of $C_T$ and $\omega$. For an increasing angular frequency $\mathcal{B}$ approaches asymptotically zero and the parameter space for bound circular orbits approaches asymptotically $\ell \in (-\infty, \infty)$ which also hints at instabilities of wormholes with ergoregion and faster rotating wormholes. 

With regard to the stability of the orbits, general statements could be made by considering static wormholes. As a necessary condition for stability it was shown that the metric functions $N$ and $K$ need to be both minimal at the throat, as their second derivatives require the same sign for stable regions of the parameter space. Moreover, if the photon orbits are stable, then also the whole spectrum of circular orbits for massive particles is stable. However, we derived a condition where even if the photon orbits are not stable, the whole spectrum of bound orbits can be stable. This condition is always trivially satisfied if the metric function $N$ is minimal at the throat. Due to the asymptotical flatness of the spacetime, this is only feasible if $g_{tt}$ possesses an outer local maximum, which can be linked to a static circular orbit at its location. Thus, unique features of the orbital properties far from the throat also influence the characteristics at the throat.

As mentioned, some of the results could be interpreted as hints to instabilities of specific wormhole solutions, which could narrow down the properties any physically feasible wormhole spacetimes should exhibit. Moreover, the presence of circular orbits at the throat can be linked to accretion disks and matter accumulations surrounding the wormhole throat, as in some fundamental accretion disk models, e. g. the thin disk or Polish doughnuts models, the accretion disks are based on the presence of stable circular orbits. The characteristics and properties of possible disk solutions could lead to different observational properties compared to disks around black holes. For example, instead of a central dark region surrounding the throat, matter accumulations could be present at the throat creating a central bright region.

\begin{acknowledgments}
P.N. gratefully acknowledges support from the Bulgarian NSF Grant KP-06-H68/7.
\end{acknowledgments}    

\bibliography{literature}

\end{document}